\documentclass[osajnl,twocolumn,showpacs,superscriptaddress,10pt]{revtex4-1} %% use 11pt for Applied Optics
\usepackage{amsmath,amssymb,graphicx}

\usepackage{hyperref}
\usepackage{breakurl}
\usepackage{amsmath}
\usepackage{graphicx}
\usepackage{amssymb}
\newcommand{\beq}{\begin{equation}}
\newcommand{\enq}{\end{equation}}

\newcommand\I{\mbox{i}}

\newcommand{\e}{\eqref}

\newcommand \bfr{{\bf r}}
\newcommand \bfk{{\bf k}}

\newcommand{\p}{\partial}

\begin{document}

%\twocolumn[ %% activate for two-column option
%%%%%%%%%%%%%%%%%% title page information %%%%%%%%%%%%%%%%%%
\title{Toward   defeating diffraction and randomness for laser beam propagation in turbulent atmosphere}

\author{Pavel M. Lushnikov and Natalia Vladimirova}
\affiliation{Department of Mathematics and Statistics, University of New Mexico, USA}
%\email{plushnik@math.unm.edu}

%% email address is required
 %\homepage{http://math.unm.edu/\textasciitilde% $\sim$ %\~{ } %\textasciitilde
 %plushnik/} %% author's URL, if desired

\begin{abstract}
A large distance propagation in turbulent atmosphere  results in  disintegration of  laser beam into  speckles. We find that
 the most intense speckle  approximately preserves    both  the  Gaussian shape and the diameter  of the initial collimated  beam   while loosing energy  during propagation.
 One per 1000 of atmospheric realizations produces at 7km distance an intense speckle above 20\% of the initial power. Such optimal
 realizations create  effective extended lenses focusing the intense speckle beyond the diffraction limit of  vacuum propagation.
  Atmospheric realizations  change every several milliseconds.   We propose to use  intense speckles to greatly increase the time-averaged power
  delivery to the target plane by triggering the pulsed laser operations only at times      of  optimal  realizations. Resulting power delivery and
  laser irradiance at the intense speckles well  exceeds   both intensity of diffraction-limited beam  and  intensity averaged over  typical  realizations.
\end{abstract}

\maketitle

\begin{figure}
\includegraphics[width=0.466\textwidth]{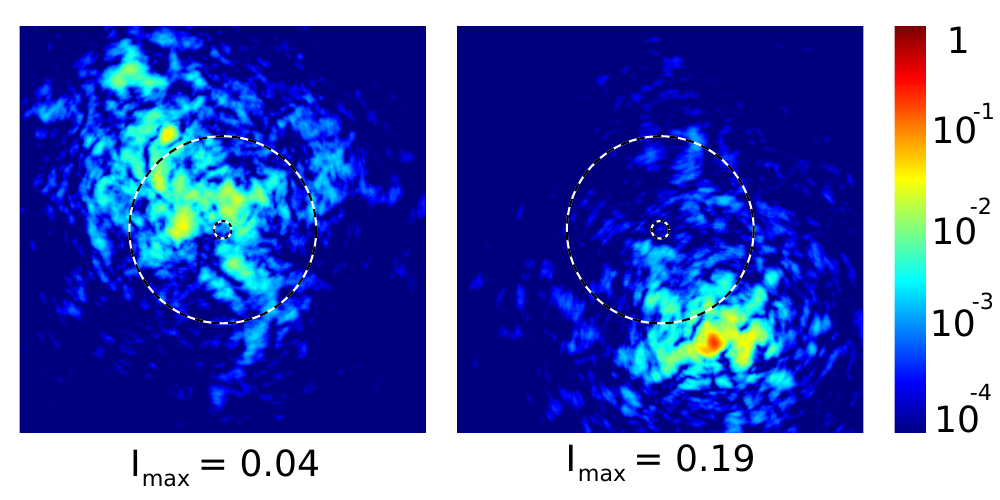}
\caption{(Color online) Distribution of laser irradiance
 $I$  in  transverse plane (only 69x69 cm central part of target screen is shown) %276.5
  after $L=7$ km propagation of the collimated Gaussian laser beam with the waist $w_0=1.5$ cm and the
   maximal intensity $I_{max}=1$ through the turbulent atmosphere in the strong scintillation regime with $\sigma_I=3.3.$
   Left panel: a typical atmospheric turbulence realization  with  $I_{max}=0.04$  ($61\%  $ of atmospheric realizations
   produce higher  $I_{max}$). Right panel: a rare realization with $I_{max}=0.19$  ($0.16\%$   realizations produce higher
    $I_{max}$).   Dashed circles show  $w_0$ and the waist of diffraction limited beam propagated in vacuum.  The initial
    Gaussian beam disintegrates into several speckles with the width of the most intense speckle being about  $w_0.$   The
    intense speckles on left and right panels carry  $4\%$ and $19\%$ of the total laser power, respectively.     }
\label{fig:7kmprofiles}
\end{figure}

Laser beam propagation though turbulent atmosphere  results in
disintegration of  laser beam into  speckles at the distances
exceeding several kilometers (strong irradiance fluctuation regime)
\cite{StrohbehnBook1978}, see Fig. \ref{fig:7kmprofiles} with
examples of such propagation. At  smaller distances (weak irradiance
fluctuation regime)  classic perturbative approaches well describe
modification of laser beam propagation due to  turbulence
\cite{TatarskiiBook1961,TatarskiiBook1971}, while
 statistically averaged  beam propagation in strong  scintillation regimes is addressed through semi-heuristic theory \cite{AndrewsPhillipsBook1998}. The strength
 of the fluctuations of the irradiance $I$ (laser beam intensity) at the target plane is characterized by the scintillation
 index  $\sigma_I\equiv \langle I^2\rangle/\langle I\rangle^2-1. $    Here and below by $\langle \ldots \rangle$ we denote an the average over the ensemble of
 atmospheric turbulence realizations. It was shown in Ref. \cite{LachinovaVorontsovJOpt2016} that a significant fraction of deviation between theoretical value of
 $\sigma_I$  \cite{AndrewsPhillipsBook1998} and simulations is due to  rare large fluctuations of laser beam intensity.
Here we study the structure of large fluctuations and propose to use them for the efficient delivery of laser energy over long distances  by triggering
the pulse laser operations only during the times of such rare fluctuations. Rear fluctuations which carry $\gtrsim19\%$  of initial power, as in Fig. \ref{fig:7kmprofiles},
occurs in $0.16\%$   realizations, and $0.1\%$ realizations carry $\gtrsim 21\%$  of initial power.  A temporal rate of change in atmospheric realizations is
affected by atmosphere conditions. In typical conditions new atmospheric realization could occur each $\sim10$ms   \cite{LachinovaVorontsovJOpt2016}.
Thus waiting for the optimal realization might take several seconds.

A  propagation of a monochromatic  beam
with a  single polarization through the turbulent media  is described by the linear
Schr\"odinger  equation (LSE) (see e.g.   \cite{TatarskiiBook1961,TatarskiiBook1971}):
\begin{eqnarray}\label{nlsdimensionall}
  \I \frac{\partial }{\partial z}\psi+\frac{1}{2k}\nabla_\perp^2\psi+k n_1({\bf r},z)
  \psi=
  0.
\end{eqnarray}
Here the beam is aligned along $z$-axis, ${\bf r}\equiv (x,y)$ are
the transverse coordinates, $\psi({\bf r},z)$ is the envelope of the
electric field,  $\nabla_\perp\equiv\left ( \frac{\partial
}{\partial x}, \frac{\partial}{\partial y}\right )$, $k=2\pi
n_0/\lambda_0$ is the wavenumber in medium, $\lambda_0$ is the
wavelength in the vacuum, $n=n_0+n_1$ is the linear index of
refraction with the average value    $n_0=\langle n\rangle$ and the
fluctuation $n_1({\bf r},z,t)$.  Time $t$ does not explicitly enters
Eq. \e{nlsdimensionall}, thus serving as parameter  distinguishing
different atmospheric realizations  so below we omit  $t$ in
arguments of all functions.

Linear absorbtion  (results in exponential decay of laser intensity
with propagation distance) is straightforward to include into Eq.
\e{nlsdimensionall}. Kerr nonlinearity can be also added to Eq.
\e{nlsdimensionall} resulting in nonlinear Schr\"odinger  Eq. which
describes the catastrophic self-focusing (collapse) for laser powers
$P$ above critical power $P_c$ ($P_c\sim3$GW for
$\lambda_0=1064\text{nm)}$
~\cite{VlasovPetrishchevTalanovRdiofiz1971,ZakharovJETP1972,LushnikovDyachenkoVladimirovaNLSloglogPRA2013}
and multiple filamentation for $P\gg P_c$
\cite{LushnikovVladimirovaOptLett2010}. At distances well below the
nonlinear length, one can consider Kerr nonlinearity as perturbation
(see e.g. Ref. \cite{VasevaFedorukRubenchikTuritsyn}) combining it
with the effect of atmospheric turbulence. Nonlinear beam combining
in atmosphere   can be also considered to fight with turbulence
\cite{LushnikovVladimirovaOptLett2014,LushnikovVladimirovaOptExpr2015}.
Such nonlinear analysis is however beyond the scope of this Letter.

We solve Eq.~\eqref{nlsdimensionall} by the standard method of
random phase screens \cite{AndrewsPhillipsBook1998} which is based
on the  approximation of statistically independent optical pulse
phase fluctuations at each screen  \cite{TatarskiiBook1971}. This
method is a version of split-step numerical method
\cite{HasegawaTappertApplPhysLett1973,FleckMorrisFeitApplPhys1976}
which separates Eq.~\eqref{nlsdimensionall} into the exactly
solvable refraction,   $\p _z\psi^R= \I kn_1({\bf r}, z) \psi^R$,
and diffraction,   $\p_z\psi^D = \frac{\I }{2k} \nabla^2_\perp
\psi^D$, parts.
 The exact solutions at the distance $\Delta z$ are given by
$ \psi^R (\bfr,z+\Delta z)  = \psi^R(\bfr,z) \exp(\I S)$ and
$\hat{\psi}^D _{k_\perp}(z+\Delta z) =\hat{\psi}^D _{k_\perp}(z) %\hat{\psi}^D_{k_\perp}(z) \exp
 %\left
 (-\I  \frac{{k_\perp}^2}{2k} \Delta z%\right
), $ respectively. Here $S\equiv k\int^{z+\Delta z}_z
n_1(\bfr,z')dz' $ is the phase shift  and
$\hat{\psi}^D_{k_\perp}\equiv (2\pi)^{-2}\int \psi(\bfr,z)e^{-\I
\bfr\cdot\bfk_\perp }d\bfr  $ is the Fourier transform (FT) for the
transverse wavevector $\bfk_\perp=(k_x,k_y).$ Sequential combining
both solutions at each step $\Delta z$ (requires  performing both FT
and inverse FT), while decreasing $\Delta z$ ensures convergence  to
the solution of  Eq.~\eqref{nlsdimensionall}.

The method of random phase screen  approximates FT of the phase
shift  at the refraction step as   $\hat{S}_{k_\perp} = \hat{\xi}
_{\bfk_\perp}k \sqrt{2\pi \hat{\Phi} _{\bfk_\perp}\Delta z }$, where
$\hat{\Phi} _{\bfk_\perp}\equiv\hat{\Phi}
_{\bfk_\perp,\kappa=0}\equiv(2\pi)^{-3}\int D(\bfr,z) e^{-\I(
\bfr\cdot\bfk_\perp+\kappa z)} |_{\kappa=0}d\bfr dz$ is the  FT over
$\rho\equiv(\bfr,z)$ of the  structure function, $D(\rho)\equiv
\langle [n_1(\bfr,z)-n_1({\bf 0},0)]^2\rangle,$ evaluated at the
zero component $\kappa=0$ of the wavevector in $z$ direction
\cite{RytovKravtsovTatarskiiBook1989}. The Kolmogorov-Obukhov law
$D(\rho)\simeq C_n^2 \rho^{2/3}$ is valid for the atmospheric
turbulence (at  $l_0\ll \rho\ll L_0)$ which implies $\hat{\Phi}
_{\bfk_\perp}= 0.033 \, C_n^2 {k_\perp}^{-11/3},$
$|\bfk_\perp|=k_\perp$       \cite{RytovKravtsovTatarskiiBook1989}.
Here $l_0$ is the inner scale of turbulence, typically a few mm, and
$L_0$ is the outer scale typically ranging from hundred meters to
kilometers.  The modification of    $\hat{\Phi} _{\bfk_\perp}$ for
both   $k_\perp\gtrsim2\pi/L_0$ and   $k_\perp\lesssim2\pi/L_0$  is
straightforward to implement
\cite{RytovKravtsovTatarskiiBook1989,AndrewsPhillipsBook1998}. We
found in agreement with Ref. \cite{LachinovaVorontsovJOpt2016},
 that   the simplest numerical
cutoff described below does not   affect the results of simulation for our range of parameters. These  parameters include the size of the square
computational domain (the transverse screen  size) $L = L_x =
L_y=276.5$ cm  with the uniformly distributed  $N \times N$ points in
that domain and $N=1024.$ It implies that $-\pi N/L\le k_x(k_y)\le
\pi N/L$  which defines the upper cut-off in $k_\perp$ variable, while the
elementary step $\Delta k=2\pi/L$ of the numerical grid $\bfk_{{\bf
j}\perp}\equiv\Delta k(j_x,j_y), \ -N/2\le j_x(j_y)\le N/2$  in
$\bfk_\perp$ determines the lower cut-off.  Also $ \hat{\xi}
_{\bfk_\perp}$ are the uncorrelated complex Gaussian random
variables on the grid  $\bfk_{{\bf j}\perp}$,
such that $\langle\hat{\xi}_{\bfk_{{\bf
j}\perp}}\rangle=\langle\hat{\xi}_{\bfk_{{\bf
j}_1\perp}}\hat{\xi}^*_{\bfk_{{\bf j}_2\perp}}\rangle=0$ for ${\bf
j}_1\ne{\bf j_2}$  and $\langle|\hat{\xi}_{\bfk_{{\bf
j}_1\perp}}|^2\rangle=(\Delta k)^{-2}.$ Here $*$ means complex
conjugation and the real values of $S$ are ensured by the condition
$\hat{\xi}_{-\bfk_{{\bf j}\perp}}=\hat{\xi}^*_{\bfk_{{\bf
j}\perp}}$.   This numerical method  is similar to
Ref.~\cite{LachinovaVorontsovJOpt2016}, except that
Ref.~\cite{LachinovaVorontsovJOpt2016} used top-hat probability density function (PDF) for
 $\hat{\xi}_{\bfk_{{\bf j}\perp}}$instead of Gaussian PDF.  We also verified that top-hat PDF produces
 essentially the same  results  (nearly visually indistinguishable on the plots below)  in comparison with
Gaussian PDF which is expected from the central limit theorem \cite{FellerBook1957}  for $N\gg 1.$

Physical parameters for our  simulations are  $\lambda = 1.064 \mu{\rm m}$, the propagation distance $z_{\rm final}=7$ km with $\Delta z =
350$m, $C_n^2 = 10^{-14} {\rm m}^{-2/3} = 4.64 \times 10^{-16} {\rm
  cm}^{-2/3}$ and a collimated input Gaussian laser  beam $\psi(\bfr,0) = \exp(- r^2/w_0^2)$ with the waist $w_0 = 1.5$~cm of
unit intensity.   Examples of simulations are shown in Fig.   \ref{fig:7kmprofiles}. The size of ensemble
is typically $4\cdot10^4$   atmospheric realizations. The averaged maximum of irradiance $\langle I_{max}\rangle=5.10433\cdot 10^{-2}$
(here $I_{max}$ is the maximum intensity in the target plane)
and the averaged irradiance $I_{\, center}=2.86788\cdot 10^{-3}$ at  the center of the target plane  with $z_{\rm final}=7$ km.
Increase of either   $C_n^2$
or $w_0$ requires decrease of $\Delta z$ to keep high numerical precision.%, although details of required

It was shown in Ref. \cite{LachinovaVorontsovJOpt2016} that  the
accurate calculation (or measurement from experiment) of $\sigma_I$
requires the ensemble of $\gtrsim 10^5$ realizations (because of
giant fluctuations of laser intensity) which is unpractical  because
atmospheric conditions are usually not stationary  at the timescale
required   for measurements of such large ensembles (hours), i.e.
the time dependence of $C_n$ becomes essential.  We argue that in
this case  $\sigma_I$ turns to be of limited usefulness because it
assumes the approximation of stationary stochastic process  which is
not valid due to the time dependence of $C_n.$ Instead, we
focus on the study of individual large fluctuations of laser intensity which qualitatively could be
interpreted as looking into optimal realizations %(instantons)
of  atmospheric turbulence through the optimal fluctuation theory. That idea was pioneered in Ref. \cite{LifshitzUspFizNauk1964}
for condensed matter, reinvented in field theory in Ref. \cite{LipatovJETP1977} and  found in many applications ranging from fluid
turbulence to \cite{FalkovichKolokolovLebedevMigdalPRE1996,ChertkovPRE1997,GrafkeGrauerSchaferJPhysA2015} to nonlinear optics
\cite{FalkovichKolokolovLebedevTuritsynPRE2001,LushnikovVladimirovaOptLett2010,ChungLushnikovPRE2011}.

\begin{figure}
\includegraphics[width=0.48\textwidth]{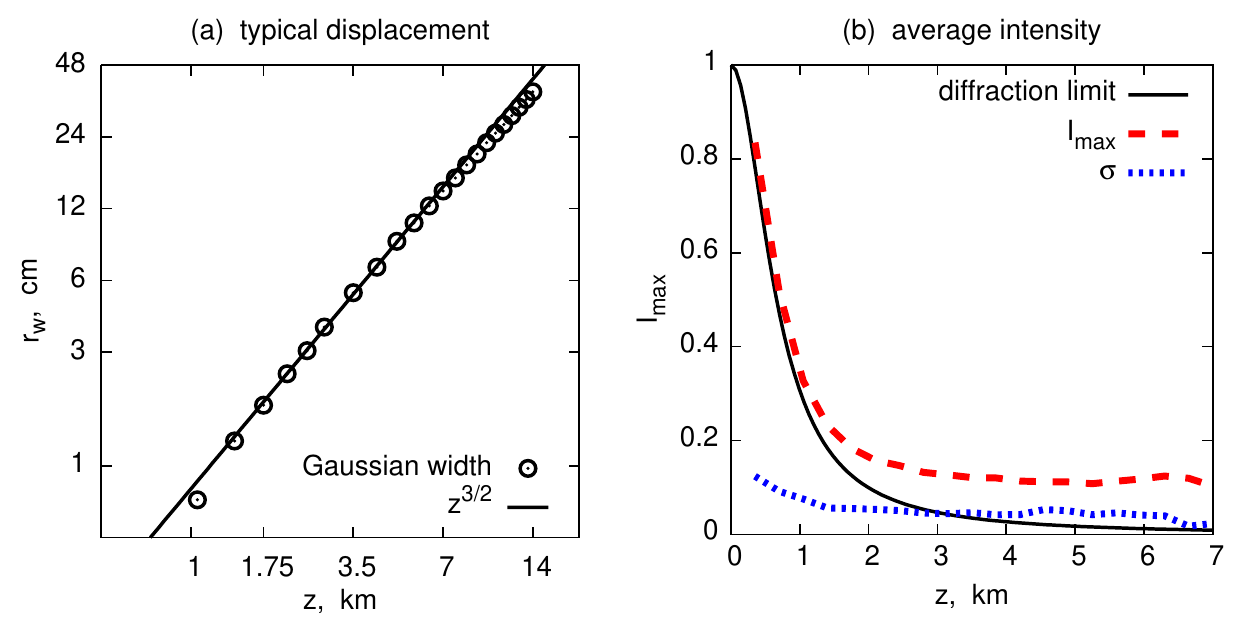}
\caption{(Color online) (a) Mean square displacement $r_w$ from
simulations (circles) vs. the accelerated random walk law $z^{3/2}$
(solid line). (b) Dashed line shows  $\langle I_{max}(z)\rangle$ for
OB averaged over 50 realizations with
$I_{max}(z=7\text{km})\simeq0.13$ Dotted line  represents  the
standard deviation from     $\langle I_{max}(z)\rangle$.  Solid line
shows   $I_{max}(z)$ for diffraction-limited beam.   }
\label{fig:Imaxz}
\end{figure}

 We identified that the optimal laser fluctuation at large propagation distance   $z\gtrsim 3$ km  is  reasonably well approximated
  by the Gaussian  beam (we called it optimal beam (OB) below) in the general approximate form
  $\psi_{optimal}(\bfr,z) = I_{max}(z)^{1/2}\exp(- [\bfr-\bfr_0(z)]^2/w(z)^2+\I [\bfr-\bfr_0(z)]^2\alpha(z)+\I\bfk_{0\perp}(z)\cdot\bfr+\I\phi(z)).$ Here
  a maximum of intensity, $I_{max}(z)^{1/2}$,  is located at the OB center,  $\bfr=\bfr_0(z),$ the OB tilt is determined by $\bfk_{0\perp}$ and the OB
  waist $w(z)$ fluctuates  with the propagation distance $z$.   Also $\phi(z)$ is the fluctuating phase shift and $\alpha(z)$ determines a fluctuating curvature of OB front.
\begin{figure}
\includegraphics[width=0.48\textwidth]{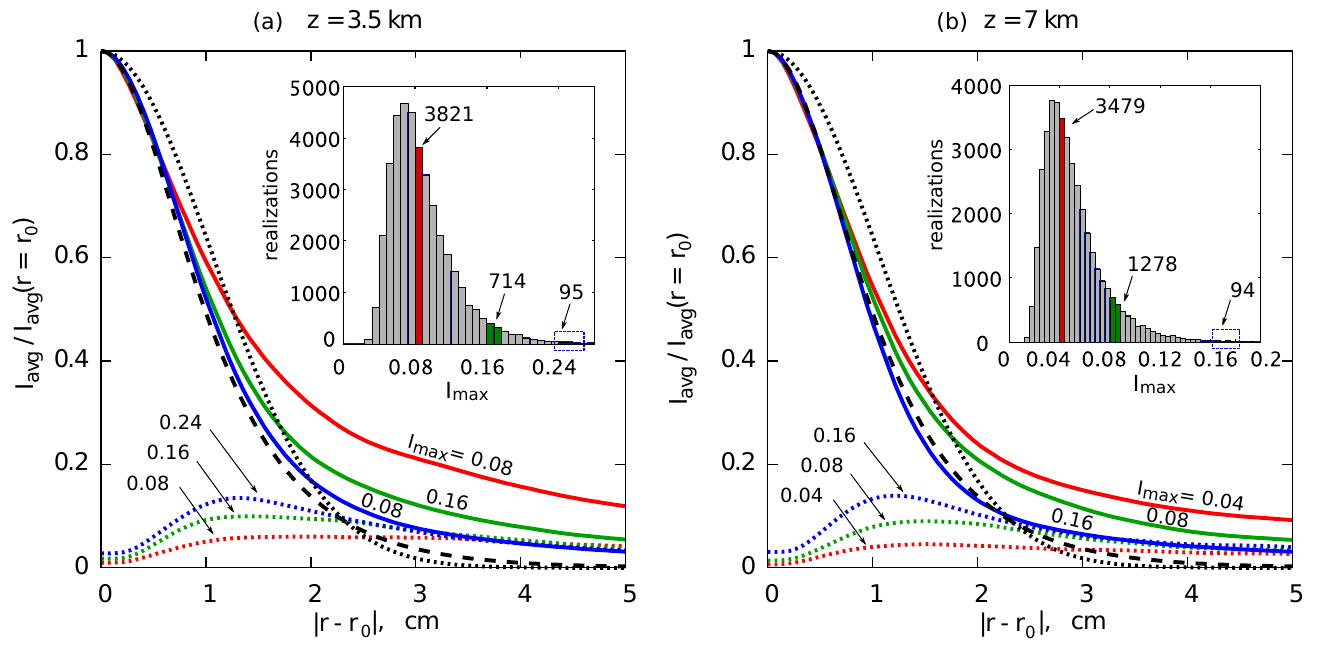}
\caption{(Color online) Averaged intensity profiles $I_{avg}(\bfr)$ of OB centered at the location $\bfr=\bfr_{0}$ of intensity maximum, $I(\bfr =\bfr_{0})=I_{max}$,
 for propagation to $z=z_{final}=3.5\, \text{km}$ and $7\,\text{km}$. Each solid line is normalized to 1 at maximum and  represents averaging  over angles  and over 50
  OB realizations with the same final value of $I_{max}(z_{final})$. The  final value is chosen within the corresponding  bins (the bin width is 0.1 of $I_{max}$) of
  the histograms for  $I_{max}(z_{final})$ realizations in the insets. The bins are selected near the maximum of PDF of   $I_{max}(z_{final}),$  in the  tail  and the
   far tail.   Total number of realizations in each histogram is $4\cdot 10^4$ (the total number of
  realizations in each beam is listed in the inset). The  standard deviation of OB profiles are shown by corresponding short-dashed lines. Dotted line
   represents the initial Gaussian beam and  thick dashed line shows  $I= 1/\cosh(2|\bfr-\bfr_{max}|/w_0)$ for comparison.    Typically, the waist of OB is $\sim 20\%$
   narrower than $w_0$.     Similar results   were also
  obtained in simulations with $w_0 = 3$
  cm.
  }
\label{fig:OBprofiles}
\end{figure}
Fig. \ref{fig:OBprofiles} shows zoom  into several  OB realizations all centered at $\bfr=\bfr_0(z)$  with amplitude rescaled to one. It is seen that   the averaged rescaled intensity profiles $I_{avg}(\bfr)/I_{avg}(\bfr=\bfr_{0})$  of these OBs are reasonably well match of the  initial beam $|\psi(\bfr,0)|$ with $w(z)\approx w_0$ and a typical deviation of $\sim20\%$.  Fig. \ref{fig:Imaxz}a  shows that  $ \bfr_0(z)$ experiences  the accelerated random walk   $ |\bfr_0(z)|\propto z^{3/2}$   due to random Gaussian fluctuations of $\bfk_{0\perp}(z)$ from $S(\bfr).$  %

Note that OB appears only after the disintegration of the initial Gaussian beam into speckles. It is seen in Fig. \ref{fig:Imaxz}b
  that at small propagation distance, $z\lesssim 1$ km (propagation in weak irradiance
fluctuation regime), the maximum irradiance $I_{max}$
%$I_{max}$
approximately follows  the diffraction limited result $I_{max,diff}=
I_0 (1+z^2/z_R^2)^{-1} $, $z_R=k w_0^2/2,$ while at larger  distance
$2\ \text{km} \lesssim z\lesssim 3$ km propagation in  moderate
irradiance fluctuation regime)  $I_{max}$
%$I_{max}$
 significantly deviates  from $I_{max,diff}$. At the same range, $2\ \text{km} \lesssim z\lesssim 3$ km, the beam disintegrates
  into speckles with the most intense speckle forming OB. At larger distances,  $ z\gtrsim 3$ km, OB amplitude fluctuates about
  approximately $z$-independent value   $\langle I_{max}(z)\rangle$  as seen in Fig. \ref{fig:Imaxz}b. According to Fig. \ref{fig:OBprofiles},
    OB waist fluctuates about $w_0$, so the optical power in OB is also approximately constant at  $ z\gtrsim 3$ km. Qualitatively we  interpret
     this behavior as random multiple focusing-defocusing events of OB at random screens which compensate the diffraction in average.

\begin{figure}
\includegraphics[width=0.48\textwidth]{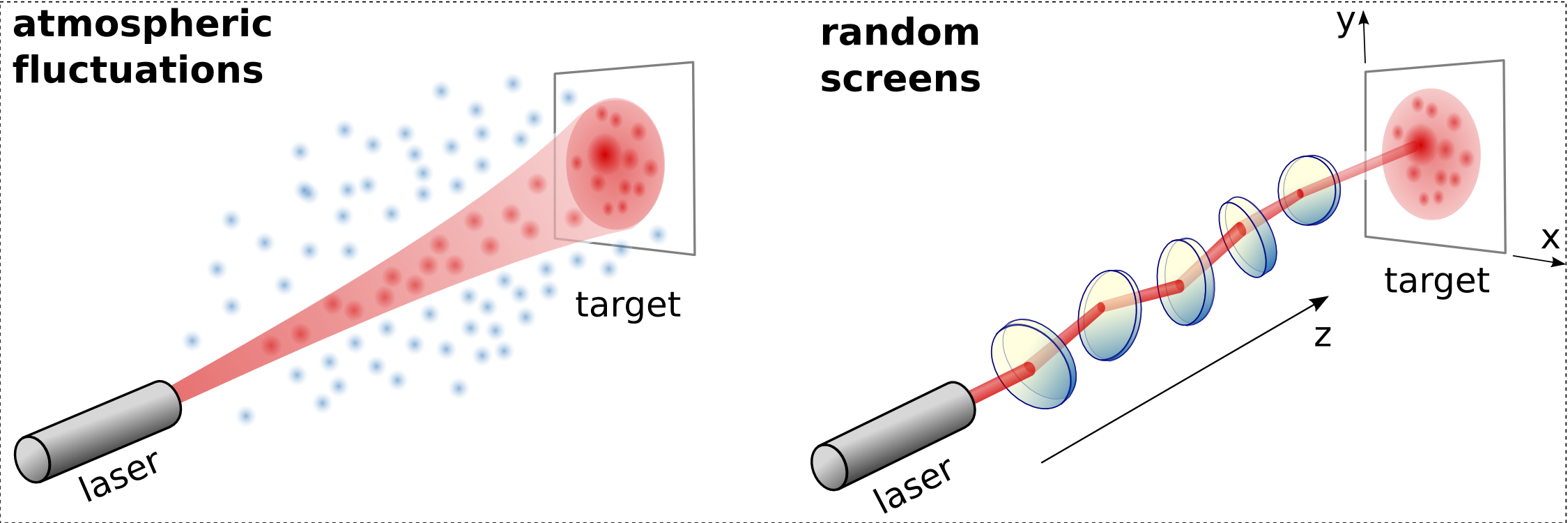}
\caption{(Color online) A sketch of laser beam propagation through
random fluctuations of refraction index in turbulent  atmosphere in
strong irradiance fluctuation regime. Left panel: entire spread of
beam is shown. Right panel: OB propagation is shown with scattering
on random fluctuations at random screens interpreted as small lenses
with random displacement, tilt and  focal lengths.}
\label{fig:7schematicslenses}
\end{figure}

To explain why the intensity profile of OB is close to Gaussian we
recall that each random phase screen modifies   $\psi$ into $\psi
e^{\I S(\bfr)}$. Neglecting the effect of small scale fluctuations
of $S(\bfr)$ on OB dynamics, we expand  $S(\bfr)$ near the center of
OB into Taylor series as ${
S(\bfr)}=S(\bfr_{0})+(\bfr-\bfr_{0})\cdot\nabla
S(\bfr_{0})+\sum^2_{l,m=1}(1/2)(x_l-x_{l,{0}})(x_m-x_{m,{0}})\nabla_l
\nabla_m S(\bfr_0)+O(|\bfr-\bfr_0|^3),$ where $(x_1,x_2)\equiv(x,y)$
and $\nabla_l\equiv\p/\p_l$. Each derivative of $S$ is the Gaussian
random variable. Then the linear term $(\bfr-\bfr_0)\cdot\nabla
S(\bfr_0)$ ensures a small random reorientation of OB about $z$
direction at each phase screen.  The quadratic form
$\sum^2_{l,m=1}(1/2)(x_l-x_{l,0})(x_m-x_{m,0})\nabla_l \nabla_m
S(\bfr_0)$ can be diagonalized by the linear transform of
$\bfr-\bfr_0$ and is responsible for the change of curvature of OB
front. Both linear and quadratic  terms can be qualitatively
interpreted  as multiple thin lenses  located in the  plane of each
phase screen as schematically shown in Fig.
\ref{fig:7schematicslenses}.
 Linear terms are responsible for the random shift of the center of lenses in transverse plane or small tilt  with respect to the
  transverse plane. Either mechanism results in the random change of slopes and wander of OB   in transverse direction as shown in Fig. \ref{fig:Imaxz}a and Fig.
\ref{fig:7schematicslenses}.  The quadratic terms can be interpreted as the  action of multiple small  focusings/defocusing lenses on
 the curvature of the OB front during propagation.

A general (non-Gaussian solution) of Eq. \e{nlsdimensionall} for
propagation between screens can be represented through the expansion
in Hermite-Gaussian modes with the Gaussian beam  being the zeroth
mode of that expansion \cite{SiegmanLasers1986}.
 $O(|\bfr-\bfr_0|^3)  $ term in $S(\bfr)$   distorts the initial Gaussian beam by producing nonzero Hermite-Gaussian modes at
 each phase screen. These  modes form ripples in $I(\bfr)$ around the main beam  so that the total optical power is conserved.
 During the initial propagation,   $z\lesssim 3 $ km,  a fraction of optical power  of these ripples that is returned to the main
 beam at each subsequent  phase screen is small. This process continues  until  OB is formed which corresponds to the approximate
 statistical steady state  for $z\gtrsim 3$ km. On these propagation distances, the small fraction of power lost from OB to higher
 Hermite-Gaussian modes at each random screen is approximately compensated  by the power returned to OB from surrounding non-small
 ripples. This effect is however small between neighboring screens, which explains why OB needs to be close to Gaussian form with $w\simeq w_0.$

In conclusion, we found that OB carry $\gtrsim21\%$ of initial power
in 1 per 1000 realizations. One can identify optimal atmospheric
realizations by  a lower power laser beam  which continuously
illuminates target plane.  When target camera/telescope detects an
optimal realization on the target,  it triggers the pulse operation
of the high-power laser.  The typical transverse displacement of OB
is $\sim10$ cm as seen in Fig. \ref{fig:Imaxz}a. If higher precision
of OB location is required, for instance for space-debris cleaning
\cite{VasevaFedorukRubenchikTuritsyn},  then one, in addition, can
continuously scan  a lower power laser beam   over angles to find
optimal realization for transverse OB location.

%{\sl\bf --Acknowledgments--}
{\it Acknowledgments.} The authors thank I.V. Kolokolov and V.V.
Lebedev for helpful discussions. This   work  was  supported by the
National Science Foundation(NSF) grant  DMS-1412140. Simulations
were performed  at the Texas Advanced Computing Center using the
Extreme Science and Engineering Discovery Environment (XSEDE),
supported by NSF Grant ACI-1053575.

%\bibliographystyle{osajnl}
%\bibliography{lushnikov,biblionls,biblio}

\end{document}